\documentstyle{l-aa}

\input {./timdefs.inp}
\input {epsf.sty}

\begin{document}
   \thesaurus{08         
              (01.01.2;
                02.02.1;  
               13.07.2;  
               08.02.3)} 

\title{The $1 - 12$ Hz QPOs and dips in GRS 1915+105: tracers of Keplerian
and viscous time scales?}

\author{S.~Trudolyubov\inst{1}, E.~Churazov\inst{2,1}, 
M.~Gilfanov\inst{2,1}}

\offprints{trudolyubov@hea.iki.rssi.ru}

\institute{Space Research Institute, Russian Academy of Sciences
Profsoyuznaya 84/32, 117810 Moscow, Russia
\and Max-Planck-Institut f\"ur Astrophysik,
Karl-Schwarzschild-Str. 1, 85740 Garching bei Munchen, Germany}

   \date{}

\maketitle
\markboth{S.Trudolyubov et al.}{Flaring activity of GRS 1915+105: 
correlated fast and slow variability in X-rays.} 

\begin{abstract}
We analyzed 9 {\it RXTE}/PCA observations of GRS 1915+105 in the flaring
state, when the hardness of the source spectrum was changing on the time
scales from few seconds up to 1000 seconds. The quasi periodic oscillations
(QPOs) with the frequency varying between  $\sim$  1 and 12 Hz are associated
with the episodes of harder source spectrum. In each observation we
found tight correlation between the duration of the hard episode and the
characteristic QPO frequency. For a half of the observations this
correlation matches the relation between the viscous time scale and the
Keplerian frequency, when both quantities are evaluated for various radii
in the radiation pressure dominated accretion disk. Assuming that the QPO
frequency is proportional to the Keplerian frequency at the boundary between
an optically thick accretion disk and a hot comptonization region, the
changes of the QPO frequency can then be understood as due to variations of
this boundary position on the viscous time scales.

\end{abstract}

\section{Introduction}
The X-ray source GRS 1915+105, one of the several galactic objects 
producing observable superluminal jets (\cite{MR94}), was discovered 
by {\it GRANAT} observatory as a transient in 1992 (Castro-Tirado, 
Brandt $\&$ Lund 1992). Long-term monitoring of GRS 1915+105 with 
{\it RXTE} revealed repetitive character of the source variability 
patterns. Basing on the results of spectral and timing analysis, bursting 
behavior of the source can be roughly reduced to the sequence of varying 
'hard' and 'soft' states qualitatively distinguished by their spectral and 
temporal properties (Belloni \etal 1997{\em a, b}; \cite{Markwardt99}; 
\cite{Muno99}). 

\begin{table}
\small
\caption{The list of {\it RXTE}/PCA observations of 
GRS 1915+105 used for the analysis. 
\label{obslog}} 
\begin{tabular}{cccc}
\hline
Obs.ID & Date, UT & Start, UT & Exp.$^a$, s\\
\hline
10408-01-01-01 & 06/04/96 & 05:39 & 5068 \\
10408-01-10-00 & 26/05/96 & 17:12 & 6128 \\
10408-01-38-00 & 07/10/96 & 05:44 & 9766 \\
10408-01-44-00 & 25/10/96 & 11:52 & 8250 \\
20402-01-01-00 & 07/11/96 & 05:42 & 6948 \\
20402-01-28-00 & 18/05/97 & 16:19 & 7349 \\
20402-01-33-00 & 18/06/97 & 12:58 & 6093 \\
20402-01-35-00 & 07/07/97 & 14:53 & 5718 \\
20402-01-59-00 & 17/12/97 & 02:10 & 8795 \\
\hline
\end{tabular}
\begin{list}{}{}
\item[$^a$]-- Deadtime corrected value of the PCA exposure
\end{list}
\end{table}

\begin{figure*}
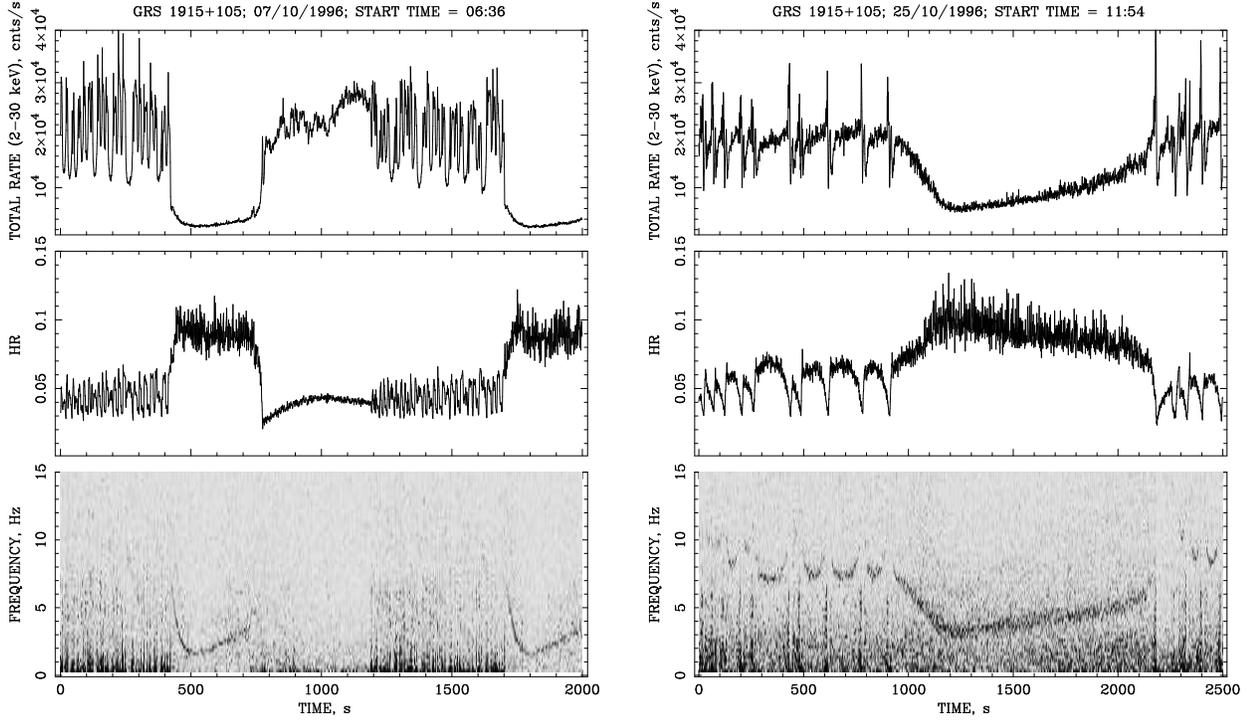

\hbox{
\epsfxsize=8.5cm
\epsffile{./Bh031.f1a}
\epsfxsize=8.5cm
\epsffile{./Bh031.f1b}
}
\caption{Part of the light curves ({\em upper panels}), corresponding 
hardness ratios ($13 - 30$ keV)/($2 - 13$ keV) ({\em middle panels}) and 
dynamic PDS ({\em lower panels}) of GRS 1915+105 for the Oct. 7, 1996 ({\it 
left panels}) and Oct. 25, 1996 ({\it right panels}) observations ($2 - 30$ 
keV energy band, PCA data). The QPO peak appears as 'U'-shaped dark band in 
the {\it lower panels}. The light curves were not corrected for the
instrument dead time which was $5 - 20 \%$. The overall count rate 
corresponds to 5 Proportional Counter Units of the PCA detector. \label{lc_pds_1}}
\end{figure*}

\begin{figure*}
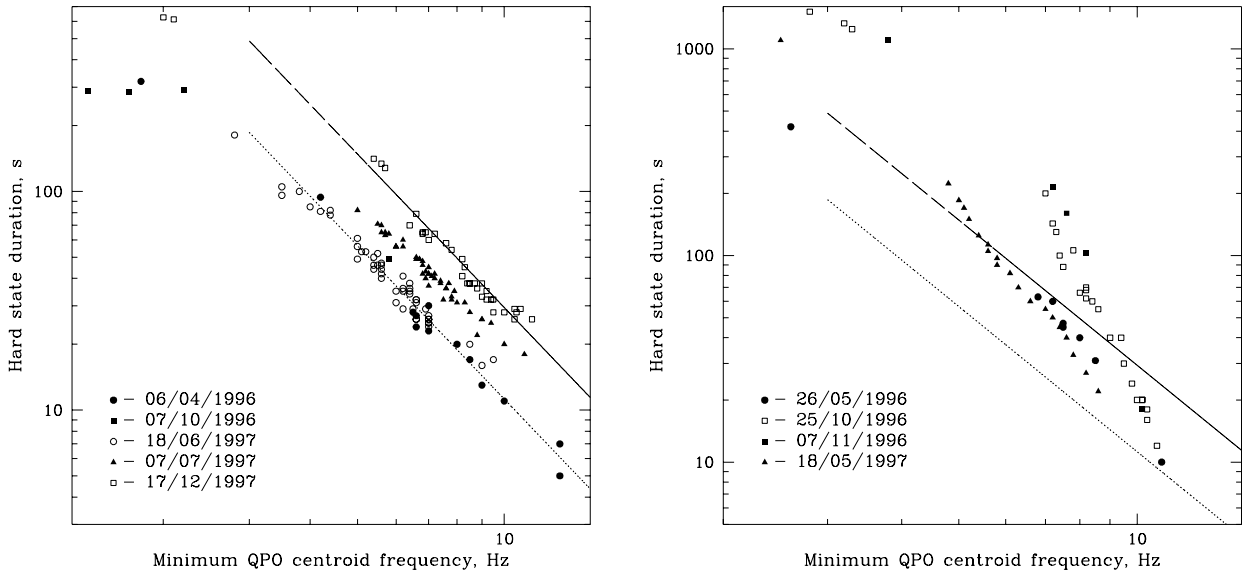

\hbox{
\epsfxsize=8.5cm
\epsffile{./Bh031.f2a}
\epsfxsize=8.5cm
\epsffile{./Bh031.f2b}
}
\caption{The relation between the duration of a hard episode and a 
corresponding minimal QPO frequency for the flaring state observations 
listed in Table 1 (first group -- {\it left panel}; second group -- 
{\it right panel}). The dependences $t_{\rm visc} \propto f_{\rm K}^{-7/3}$ of 
the viscous time scale upon the Keplerian frequency at the inner edge 
of the radiation pressure dominated disk with the mass accretion rate 
$\dot{m} \sim 0.11 (\alpha / 0.1)^{-1/2} (m/ 33)^{-2/3}$ and $\dot{m} 
\sim 0.17 (\alpha / 0.1)^{-1/2} (m/ 33)^{-2/3}$ are shown in the {\it 
left and right panels} by long-dashed and dotted lines respectively 
(see text).
\label{qpo_dip_1}}
\end{figure*}

\section{Observations and data analysis}
The observations used for the analysis are listed in Table 1. For the 
timing analysis in the $2 - 30$ keV energy range the {\it RXTE}/PCA 
(Bradt, Swank $\&$ Rothschild 1993) data in the 
'binned' and 'event' modes containing X-ray events below and above 
13 keV respectively were used. We generated power density spectra (PDS) 
in the $0.25 - 150$ Hz frequency range accumulated in the 4 s time 
intervals (Leahy \etal 1983). The PDS were linearly 
rebinned into 0.25 Hz bins. The white-noise level due to Poissonian 
statistics corrected for the dead-time effects was subtracted. For the 
determination of the QPO centroid frequencies the PDS were fitted to the 
analytic model consisting of a Lorentzian profile and a power law 
continuum in the $0.5 - 15$ Hz frequency range (typical error of the 
derived value of the QPO frequency is $\sim 0.2 - 0.3$ Hz).

\section{Results}
Two typical examples of GRS 1915+105 X-ray flux histories in the $2 - 30$ 
keV band are presented in Fig. \ref{lc_pds_1} along with hardness ratios 
($13 - 30$ keV to $2 -13$ keV) and dynamic PDS integrated over 4 s time 
intervals. The source behavior in the flaring state is characterized by
frequently occurring episodes of hard energy spectrum lasting typically 
$10 - 1000$ s. The prominent, relatively narrow variable-frequency $1 - 12$ 
Hz QPO peak (dark 'U'-shaped band in the {\it lower panels} in Fig. 1) 
is a generic feature of the source power density spectrum during hard 
episodes (\cite{Markwardt99}; \cite{Muno99}).

Using the data from our sample of flaring state observations (Table 1), we 
studied the relation between the duration of a hard episode, $t_{hard}$ 
(determined using the hardness ratio) and the lowest value of the QPO 
frequency reached (determined from the fitting of the PDS to the 
analytical model). In Fig. \ref{qpo_dip_1} the duration of a hard 
episode is plotted as a function of the minimal QPO frequency. For all 
observations used in the analysis there is a strong anticorrelation between 
these two quantities. The dependence has a nearly power law form for 
durations of a hard episode shorter than $\sim 100 - 200$ s and probably 
flattens for longer events. According to the value of the 
slope $\Gamma$ of this power law dependence the observations can be 
subdivided into two groups: $\Gamma \sim 2.1 - 2.4$ for the 06/04/96, 
07/10/96, 18/06/97, 07/07/97 and 17/12/97 observations (the first group, 
{\it left panel} in Fig. \ref{qpo_dip_1}) and $\Gamma \geq 3.0$ for the 
26/05/96, 25/10/96, 07/11/96 and 18/05/97 observations (the second group, 
{\it right panel} in Fig. \ref{qpo_dip_1}). The analysis of the flux 
histories showed that these two groups have qualitatively different 
lightcurves: for the first group the transition from 'hard' to 'soft' 
state is characterized by the growth of the $2 - 30$ keV flux (Figure 
\ref{lc_pds_1}, {\it upper left panel}), while for the second group 
this transition corresponds to the decrease of the X-ray flux (Figure 
\ref{lc_pds_1},{\it upper right panel}). In the subsequent paper 
(\cite{Trudolyubov99.2}) it will be shown that these two groups are also 
distinguished by the properties of their energy spectra: for the first 
group the soft thermal component makes a significant contribution 
to the overall luminosity in the $3 - 20$ keV energy range, for the 
second group the energy spectrum is dominated by the hard component (the 
contribution of the soft thermal component to the total luminosity is $<25
\%$).

\section{Discussion}
The study of the spectral and temporal evolution of GRS 1915+105 during the
periods of flaring activity has shown that bursting behavior of the source 
can be reduced to a sequence of varying hard and soft episodes. The hard 
episodes are distinguished by the presence of the prominent QPO peak in the 
source power density spectrum. Combined time-resolved spectral and timing
analyses revealed (i) a strong correlation between the QPO frequency 
and parameters of the soft component of the energy spectrum (presumably 
emitted by the optically thick part of the disk) on a wide range of time 
scales (Trudolyubov \etal 1999{\it a}; Markwardt \etal 1999; Muno \etal 
1999); (ii) the dependence of the duration of the hard episode upon the  
corresponding characteristic radius of the optically thick disk (derived 
through the spectral fitting with multicolor disk black body model) 
$t_{\rm hard} \propto r^{7/2}$ is similar to that expected if the duration 
of the hard episode is proportional to the viscous time scale of the radiation
pressure dominated disk (Belloni \etal 1997{\it b}). Given (i) and (ii) 
some sort of correlation between the QPO frequency and the duration of 
the hard episode is naturally expected.

To explain the observational properties of black hole and neutron star 
binaries in the {\it hard} state a number of models involving the hot 
comptonization region near the compact object surrounded by the optically 
thick accretion disk was proposed. It is often assumed that the QPO 
phenomenon is caused by interaction between these two distinct parts of 
the accretion flow occurring on the local dynamical time scale at the
boundary of these regions (\cite{msc96}; \cite{tlm98}). In the following
analysis we will also assume that the QPO frequency is proportional to 
the Keplerian frequency at the inner boundary of the accretion disk.

\begin{figure}
\epsfxsize=9.0cm
\epsffile{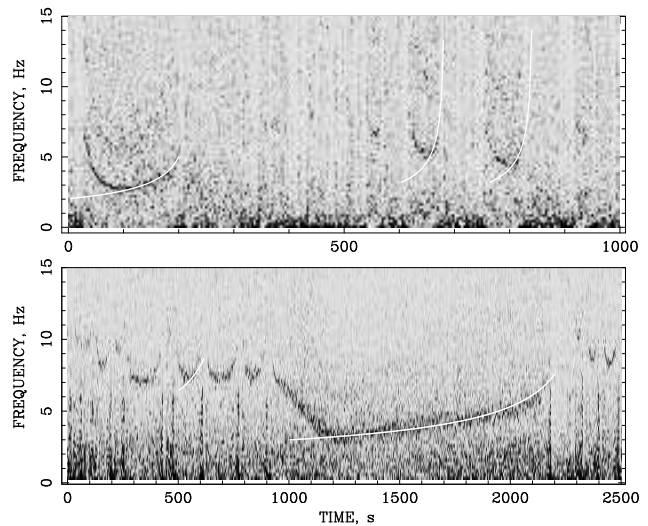}
\caption{Dynamic power density spectra for two representative observations 
(18/06/1997 -- first group ({\em upper panel}, 25/10/1996 -- second group 
{\em lower panel}). The expected time dependence of the QPO frequency 
implied by viscous evolution of the radiation pressure dominated part of 
the accretion disk is presented by a white solid lines superimposed on the 
observed QPO track. \label{obs_theory}}
\end{figure}

In the standard accretion disk theory (Shakura $\&$ Sunyaev 1973) the 
viscous time of the radiation pressure dominated disk is $t_{\rm visc} 
\sim 1.2 \times 10^{-5} \alpha^{-1} m \dot{m}^{-2} r^{7/2} \;\;s$, where 
$\alpha$ -- viscosity parameter, $m$ -- mass of the compact object in 
solar masses, $\dot{m}$ -- disk accretion rate in units of critical 
Eddington rate, $r$ -- distance to the compact object in units of 3 
gravitational radii. Introducing the local Keplerian frequency, $f_{\rm K} 
\approx 2200 \; m^{-1} r^{-3/2}$ Hz, we obtain: 
\begin{eqnarray}
t_{\rm visc} \sim 740 \alpha^{-1} m^{-4/3} \dot{m}^{-2} 
f_{\rm K}^{-7/3} \; \; s 
\end{eqnarray}

This dependence, $t_{\rm visc} \propto f_{\rm K}^{-7/3}$, reproduces well 
the relation between the duration of a hard episode and the associated minimal 
QPO frequency for the first group of observations (Fig. \ref{qpo_dip_1}, 
{\it left panel}). Moreover, for several observations from this group 
the observed relation between the duration of a hard episode and the maximal 
value of the characteristic inner radius of the disk, determined via 
spectral fitting, matches remarkably well the expected dependence of the 
viscous time scale on the radius for the radiation pressure dominated 
disk (\cite{Belloni2}). 

Assuming that the QPO frequency is proportional to the Keplerian frequency
($f_{\rm QPO} = f_{\rm K}/ l$) and duration of the hard episode is 
proportional to the viscous time ($t_{\rm hard} = t_{\rm visc}/ k$), one 
can estimate the required mass accretion rate $\dot{m}$ using equation (1):
\begin{eqnarray}
\dot{m} \sim \left ( \frac{\alpha}{0.1} \right )^{-1/2} \left 
( \frac{m}{33} \right )^{-2/3} \left ( \frac{k \; t_{\rm hard}} {100 s} 
\right )^{-1/2} (l f_{\rm QPO})^{-7/6}   
\end{eqnarray} 
In Eq. (2) compact object mass was normalized to the value of 33 
M$_{\odot}$, implied by the interpretation of the 67 Hz QPO feature 
as a signature of a Keplerian oscillations near the last marginally stable 
orbit around a Schwarzschild black hole (\cite{Morgan97}). The inferred 
values of the accretion rate for the observations from the first group 
are in the range $\dot{m} \sim (0.11 - 0.17) (\alpha / 0.1)^{-1/2} 
(m/ 33)^{-2/3} k^{-1/2} l^{-7/6}$ (Fig. \ref{qpo_dip_1}). This conclusion 
is supported by the results of the spectral analysis (Trudolyubov \etal 
1999{\it b}) where the bolometric luminosity of the soft spectral component 
corresponding to a given value of QPO frequency was used as a measure of 
the mass accretion rate.

If the QPO frequency is associated with the selected region in the accretion
disk (e.g. inner edge of the optically thick disk) which is moving radially
on the viscous time scale, then we expect correlation of the QPO frequency 
and the rate of the frequency change. Assuming that rise of the
QPO frequency is caused by the inward motion of the inner edge of the
radiation pressure dominated accretion disk one can write: 
\begin{eqnarray}
\left ( \frac{df_{\rm QPO}}{dt} \right ) \; = \; \left 
( \frac{df_{\rm QPO}}{dr} \right ) v_{\rm r}(r) \sim \alpha \dot{m}^{2} 
m^{7/3} f_{\rm QPO}^{10/3},
\end{eqnarray}
where $v_{\rm r}(r)$ -- radial velocity of the matter in the disk at radius 
$r$ (Shakura $\&$ Sunyaev 1973). Integrating Eq. (3), we obtain:
\begin{eqnarray}
f_{\rm QPO}^{-7/3}(t_{0}) - f_{\rm QPO}^{-7/3}(t) \; = \; A (t - t_{\rm 0}), 
\end{eqnarray}
where $A \sim \alpha \dot{m}^{2} m^{7/3}$. In Fig. \ref{obs_theory} this 
dependence is shown in comparison with the observed QPO evolution for 
two observations from the first (18/06/1997) and the second (25/10/1996) 
groups. For the first group both initial decrease and 
following rise of the QPO frequency during the hard episodes are generally 
described by Eq. (4) (Fig. \ref{obs_theory}, {\it upper panel}) but with 
different values of coefficient $A$. For the rise phase the value of 
coefficient $A$ remains practically the same on a time scale of an 
individual observation ($\sim 10^{4} \; s$). For the second group of 
observations only rise of the QPO frequency is generally described by 
Eq. (4), while the initial decay often has a more complicated structure 
(Fig. \ref{obs_theory}, {\it lower panel}). In addition, an extended 
plateau in the time dependence of the QPO frequency near its minimum is 
sometimes present. This might explain the deviation of the dependence of 
$t_{\rm hard}$ on the QPO frequency from the $t_{\rm hard} \propto f_{\rm QPO}^
{-7/3}$ law for the second group of observations.

\section{Conclusions}
We analyzed the temporal behavior of GRS 1915+105 using the set of 9 {\it 
RXTE}/PCA observations representing typical patterns of the source X-ray
evolution during the periods of flaring activity.\\
1. For all individual observations the tight correlation between the 
duration of a hard episode and the minimal QPO frequency was found.\\
2. For $\sim$half of the observations this correlation is satisfactory
described by the standard radiation pressure dominated disk theory assuming
that the QPO frequency and the duration of a hard episode are proportional 
to the Keplerian frequency and the viscous time on the inner edge of the disk 
respectively. The derived values of the mass accretion rate for these 
observations are in the range $\dot{m} \sim (0.1 - 0.2) (\alpha / 0.1)^
{-1/2} (m/ 33)^{-2/3}$. Not all observations fit this dependence. In fact 
for the large fraction the relation $t_{\rm hard} \propto f_{\rm QPO}^{-7/3}$ 
breaks. It should be noted, that for these observations the correlation 
$t_{\rm hard} \propto r^{7/2}$ (Belloni \etal 1997{\it b}) also fails. \\
3. The temporal evolution of the QPO frequency during the rise phase of 
the hard episodes can be explained as a change of the Keplerian frequency 
at the inner edge of optically thick accretion disk caused by its viscous 
motion. 

{\it Acknowledgments} This research has made use of data obtained through 
the HEASARC Online Service, provided by the NASA/GSFC. S. Trudolyubov is 
partially supported by RBRF grants 96-15-96343, 97-02-16264 and INTAS 
grant 93-3364-ext.

\end{document}